\documentstyle[amssymb,aps,epsfig,float,twocolumn]{revtex}
\restylefloat{figure}
\title{Magnetic order in the quasi-two-dimensional easy-plane XXZ model}
\author{D.~Ihle}
\address{Institut f\"ur Theoretische Physik, Universit\"at Leipzig,
  D-04109 Leipzig, Germany} 
\author{ C. Schindelin}
\address{COI GmbH, Erlanger Strasse 62, D-91074 Herzogenaurach}
\author{H.~Fehske}
\address{Physikalisches Institut, Universit\"at Bayreuth, 
  D-95440 Bayreuth, Germany\\{\rm (\today)}}
\address{~\parbox{14cm}{\rm
    \medskip
A Green's-function theory of antiferromagnetic short-range and long-range
order (LRO) in the $S=1/2$ 
quasi-two-dimensional easy-plane XXZ model is presented.
As the main new result, {\it two} phase transitions due to
the combined influence of spatial and spin anisotropy are found, 
where below the higher and lower N\'{e}el temperature there 
occurs LRO in the transverse and in both the transverse and longitudinal
spin correlators, respectively. Comparing the theory with neutron-scattering
data for the correlation length of $\rm La_2CuO_4$, a very good agreement
in the whole temperature dependence is obtained. 
Moreover, for $\rm La_2CuO_4$, $\rm Sr_2CuO_2Cl_2$, 
and $\rm Ca_{0.85}Sr_{0.15}CuO_2$ the second phase with longitudinal LRO
is predicted to appear far below room temperature. 
 }}
\begin{document}
\maketitle
\section{Introduction}
To understand the unconventional behavior of high-T$_c$ superconductors,
which is mainly ascribed to a strong antiferromagnetic (AFM) short-range
order (SRO), the magnetic properties of the quasi-two-dimensional (2D)
parent compounds  were probed preferably by neutron scattering and NMR 
experiments, e.g., on $\rm La_2CuO_4$~\cite{Keea92a,Keea92b,MMRB97,Biea99},
 $\rm Ca_{0.85}Sr_{0.15}CuO_2 $~\cite{MRB99}, $\rm YBa_2Cu_3O_{6+x}$ 
($x\lesssim 0.4$)~\cite{MNIY93} 
and $\rm L_2CuO_4$ (L=Nd, Pr)~\cite{Maea90}. 
In particular, the staggered 
magnetization~\cite{Keea92a,Keea92b,MMRB97,MRB99,MNIY93} and
the AFM correlation length~\cite{Keea92b,Biea99} were investigated.
To analyze the experimental data, quasi-2D Heisenberg models including
a weak spin anisotropy (easy-plane XXZ models) were 
considered~\cite{Keea92a,Keea92b,MMRB97,Biea99,MRB99,MNIY93,Maea90} 
and treated by linear spin-wave theory~\cite{MMRB97,MRB99,MNIY93}
and a Schwinger-boson approach~\cite{Keea92a}. Recently, a quasi-2D 
anisotropic  (easy-axis) Heisenberg model was studied within a boson-fermion
mean-field theory~\cite{IKK99}. However, those auxiliary-field approaches
and most of the spin-wave theories for related spin models (for references
to most of the early work, see Ref.~\cite{SIH00}) yield reasonable 
results only at sufficiently low temperatures, since the temperature-dependent
SRO is not adequately taken into account. To provide a good description
of magnetic SRO at arbitrary temperatures, a Green's-function theory
based on the projection method was developed for the spatially
anisotropic Heisenberg models~\cite{SIH00,WI97,ISWF99} and for
the 1D~\cite{SFBI00} and 2D easy-plane XXZ models~\cite{FSWBI00}.

In this paper we extend our previous work to the $S=1/2$ quasi-2D 
easy-plane XXZ model
\begin{eqnarray}
  {\cal H} &=& \frac{J}{2}\,\Big[
\sum_{\langle i,j\rangle_{x,y}}\Big( S^+_i S^-_j 
+\Delta \, S^z_i S^z_j\Big)\nonumber\\&& + R_z\,
\sum_{\langle i,j\rangle_{z}}\Big( S^+_i S^-_j 
+\Delta \, S^z_i S^z_j\Big)
\Big]\,.
\label{xxzmod}
\end{eqnarray}
Here, $\langle i,j \rangle_{xy}$ and $\langle i,j \rangle_{z}$
denote nearest-neighbor (NN) sites in the xy plane and along the z direction
of a simple cubic lattice, respectively, $R_z=J_z/J<1$ describes the AFM
interplane coupling (throughout we set $J=1$), and the spin anisotropy 
parameter $\Delta$ is considered in the AFM region $0<\Delta\leq 1$.
In the limiting cases $\Delta =1$ and $R_z=0$ the model~(\ref{xxzmod})
reduces to the models studied in Refs.~\cite{SIH00} and~\cite{FSWBI00}, 
respectively.

Our Green's-function theory outlined in the Appendix 
is based on an approximate time evolution of spin operators,
e.g., $ -\ddot{S}^z_{\bf q}=(\omega_{\bf q}^{zz})^2 S^z_{\bf q}$,
resulting from a decoupling of three-spin operator products
which is improved by the introduction of vertex parameters. 
For the temperature dependence of some vertex parameters assumptions 
are made which are detailed and motivated in the Appendix.
Within this theory  
we examine the combined effects of spatial and spin anisotropy on 
the AFM long-range order (LRO) at $T=0$ (Sec.~II), 
the N\'{e}el transition temperature, and on the AFM 
correlation length (Sec.~III). In Sec.~IV we compare our results
with experiments.  For $\rm La_2CuO_4$, 
$\rm Sr_2CuO_2Cl_2$, and $\rm Ca_{0.85}Sr_{0.15}CuO_2$  
our theory predicts, in addition to the usual N\'{e}el transition,
a further transition far below room temperature, 
where the spin correlators between the z-components start to develop
AFM LRO.
\section{Ground-state long-range order}
Analyzing the magnetic LRO described by the staggered magnetizations
$m^{\nu}$ [$\nu=\pm, zz$; cf. Eq.~(\ref{m})] at $T=0$ as functions  
of $R_z$ and $\Delta$, we obtain transverse LRO in the whole parameter
region considered and two solutions differing in the existence of
longitudinal LRO. That is, we obtain a phase with $m^{zz}=0$
(phase~I) and a phase with $m^{zz}\neq 0$ (phase~II), where in both 
phases we have $m^{+-}\neq 0$. The stabilization of longitudinal 
LRO in the easy-plane region by the interplane coupling 
may be due to the reduction of quantum spin fluctuations 
in higher dimensions. 

\begin{figure}[!htb] 
\epsfig{file= 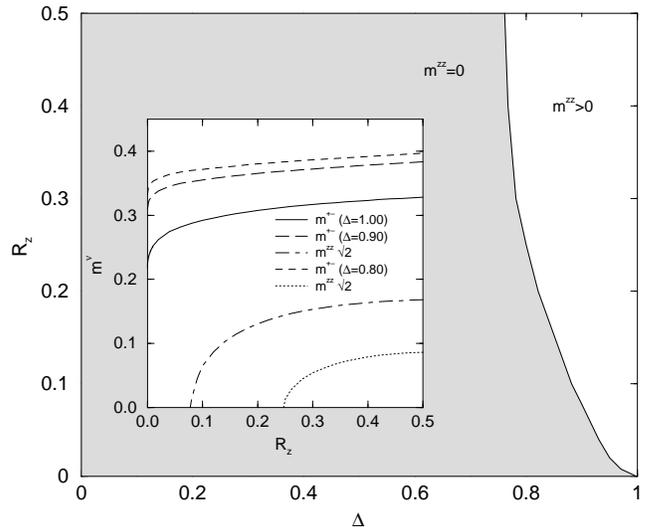, width =.98 \linewidth}
\caption{$R_z-\Delta$ phase diagram and transverse and longitudinal
zero-temperature magnetizations (inset) in the quasi-2D easy-plane
XXZ model.}
\label{fig1}
\end{figure}

In Fig.~\ref{fig1} the $R_z-\Delta$ phase diagram is shown, 
where for $R_z\neq 0$ the transition across the 
phase boundary denoted respectively by $R_{z,c}(\Delta)$
and $\Delta_c(R_z)$ is found to be of second order (cf. inset).
Let us consider the phase transition in the vicinity of the 
critical point $(\Delta,R_z)=(1,0)$ in more detail. In our
approach the solution for $m^{zz}$ turns out to depend sensitively
on the input data for $\partial e(\Delta,1)/\partial \Delta \equiv e'$
used to determine the vertex parameter $\alpha_2^{zz}$,
where $e(\Delta,1)$ denotes the ground-state energy of the 
2D XXZ model (see Appendix). Taking for $e(\Delta,1)$ the 
exact diagonalization (ED) data on lattices with up to 36
sites (without finite-size scaling) from Ref.~\cite{FSWBI00},
we obtain $\lim_{R_z\to 0} \Delta_c(R_z)\equiv\Delta_0=0.958$.
On the other hand, taking the Monte Carlo (MC) data from
Ref.~\cite{OK88}, which have to be interpolated between the few
available points $\Delta = 0, \pm 0.5, \pm 1$, we get 
$\lim_{\Delta \to 1} R_{z,c}(\Delta)\equiv R_{z,0}=4.08\times 10^{-2}$.
However, as is well known, at $R_z=0$ there is no
longitudinal LRO for $0<\Delta<1$ which means that we must have
$\Delta_0=1$. Moreover, we expect the interplane coupling to be a relevant
perturbation with respect to the stabilization of LRO analogous 
to the situation in the 2D spatially anisotropic Heisenberg 
model~\cite{ISWF99,San00}, where LRO at a finite arbitrary small
interchain coupling was found~\cite{San00}. Therefore, we make the 
reasonable assumption $R_{z,0}=0$. To fulfill  the requirements
$\Delta_0=1$ and  $R_{z,0}=0$ simultaneously, so that the phase boundary
touches the critical point $(\Delta,R_z)=(1,0)$, as input for
$e'$ we use a linear combination of the ED~\cite{FSWBI00} and
Monte Carlo data~\cite{OK88}, $e'=xe'_{ED}+(1-x)e'_{MC}$.
We find that the above requirements may be fulfilled,
if $x$ is chosen as x=0.44.  

In the limit $\Delta =1$ we have rotational symmetry
($C_{\bf r}^{zz}=C_{\bf r}^{+-}/2$), so that 
$\sqrt{2} m^{zz}(R_{z})=m^{+-}(R_{z})\equiv \sqrt{2/3} m(R_{z})$
with $m$ defined as in Refs.~\cite{SIH00,WI97,ISWF99}. 
In this limit, our result for $m^{+-}(R_{z})$ agrees with 
that of Ref.~\cite{SIH00}. As can be seen in the inset,
the effects of spin anisotropy on the longitudinal and transverse
LRO are opposite: We have $\partial m^{zz}/\partial \Delta >0$,
whereas $\partial m^{+-}/\partial \Delta <0$ which agrees, at $R_z=0$, 
with the Monte Carlo data~\cite{OK88} and the results of Ref.~\cite{FSWBI00}.
\section{Finite-temperature properties}
Considering the AFM LRO in the phases~I and~II
[$m^{\nu}(T=0)\neq 0$ at $R_z>R_{z,c}$] at nonzero temperatures,
the solution of the self-consistency equations~(\ref{cn}), supplemented
by the conditions for the vertex parameters [cf. Eqs.~(\ref{ellbed}) 
and~(\ref{r1}) to~(\ref{r3})], results in two second-order phase
transitions at $T_N^{+-}(R_z,\Delta)$ and $T_N^{zz}(R_z,\Delta)$
[$m^{\nu}(T_N^{\nu})=0$] with $T_N^{+-} >T_N^{zz}$ and $T_N^{zz}(R_{z,c})=0$.
\begin{figure}[!b] 
\epsfig{file= 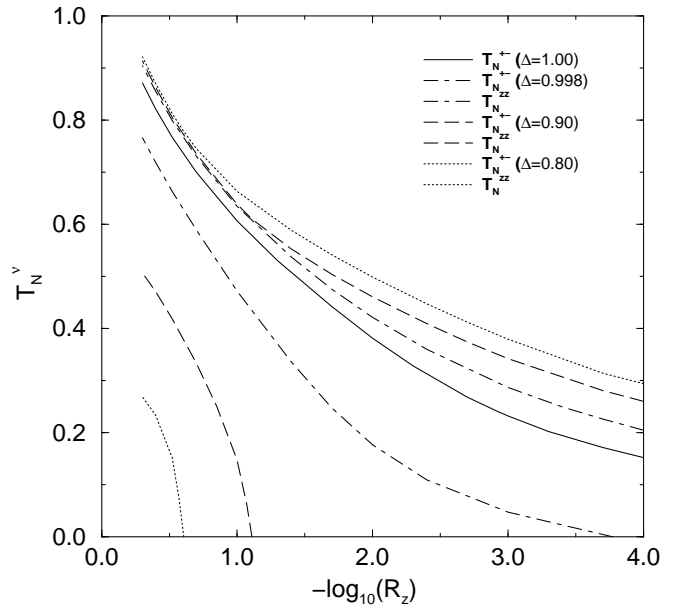, width =1.0 \linewidth}
\caption{$T-R_z$ phase diagram in the quasi-2D easy-plane
XXZ model. Below the N\'{e}el temperature $T_N^{zz}$ the phase
with longitudinal long-range order becomes stable.
The curves below the solid line belong to $T_N^{zz}$.}
\label{fig2}
\end{figure}
Figure~\ref{fig2} shows the N\'{e}el temperatures as functions of $R_z$, i.e.,
the $T-R_z$ phase diagram for different spin anisotropies.
For $R_z=0$ we obtain $T_N^{+-} =0$, in agreement with the Mermin-Wagner 
theorem. At $\Delta =1$ we have $T_N^{+-} =T_N^{zz}\equiv T_N$. As compared
with the results of Ref.~\cite{SIH00}, our values for $T_N$ are
somewhat higher (by about 9\%) due to numerical uncertainties.
On the other hand, in comparison with previous RPA and mean-field
approaches (cf. Ref.~\cite{SIH00}) our N\'{e}el temperatures are
reduced by the improved  description of SRO. For example, the
$T_N$ values found by the Schwinger-boson approach of Ref.~\cite{Ko92}
exceed our results by a factor of about~1.7 on the average.

Concerning the influence of spin anisotropy on the N\'{e}el
transitions, we obtain  $\partial T_N^{+-}/\partial \Delta < 0$
and $\partial T_N^{zz}/\partial \Delta >0$, corresponding to
the $\Delta$ dependence of $m^{\nu}$ (cf. inset of Fig.~\ref{fig1}). 
The dependence on $\Delta$ of $T_N^{+-}$ is
in qualitative agreement with the behavior found in previous 
approaches~\cite{Keea92a,Keea92b}. There, $T_N$
(being identified with $T_N^{+-}$) is given as 
$T_N/J=-2\pi M_0\{\ln |4\alpha_{eff}/
[\pi^2M_0 \ln (4\alpha_{eff}/\pi)]|\}^{-1}$ (Ref.~\cite{Keea92a})
with $M_0=0.3$, $\alpha_{eff}=4\alpha_{xy}+2R_z$, and
$\alpha_{xy}=1-\Delta$ or as $T_N/J=-4\pi\rho_S(\ln \alpha_{eff})^{-1}$
(Ref.~\cite{Keea92b}), where $\rho_S$ is the spin stiffness. 
However, in contrast to those mean-field (Schwinger boson) results,
in our theory the combined influence of spatial and spin anisotropy on
$T_N^{+-}$ cannot be expressed in terms of a single effective parameter.
Considering the variations $\delta R_z$ and  $\delta \alpha_{xy}$
under the condition  $\delta T_N^{+-}=0$ we get ${\rm sgn} 
(\delta\alpha_{xy} )= -{\rm sgn} (\delta R_z)$.  
Whereas the $T_N$ formulas quoted above yield $\delta\alpha_{xy}=
-\delta R_z/2$ for all $R_z$, from Fig.~\ref{fig2} we obtain a
$R_z$ dependent relation between $\delta\alpha_{xy}$ and  $\delta R_z$.
\begin{figure}[!b] 
\epsfig{file= 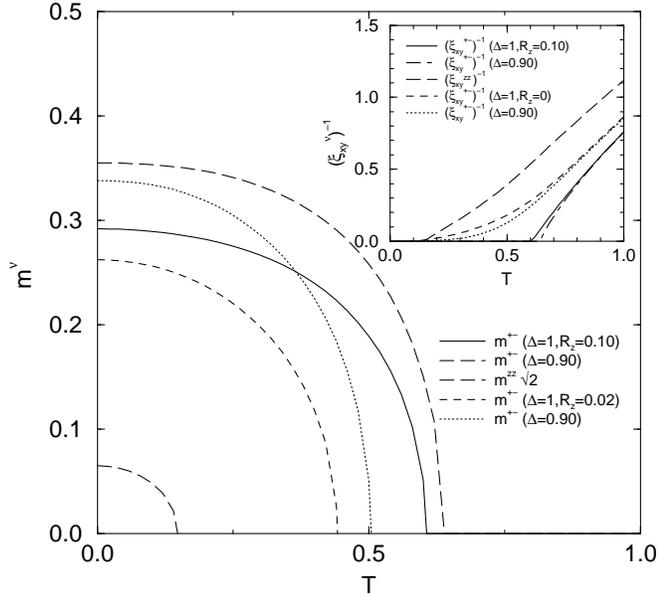, width =1.0 \linewidth}
\caption{Staggered magnetizations versus temperature.
The inset shows the inverse intraplane correlation lengths
above the corresponding N\'{e}el temperatures.}
\label{fig3}
\end{figure}

In Fig.~\ref{fig3}, our numerical results for the temperature dependence of
the magnetization $m^{\nu}$  are depicted. They
may be described by a $T^2$ decrease at low enough 
temperatures (3D behavior) and by
$m^{\nu}(T)\propto (1-T/T_N^{\nu})^{1/2}$ for temperatures
close to $T_N^{\nu}$, as was also found in the $\Delta =1$ 
limit~\cite{SIH00}. The influence of spatial and spin anisotropy
on $m^{\nu}(T)$ is analogous to that on $m^{\nu}(T=0)$ shown
in the inset of Fig.~\ref{fig1}.
 
In the inset of Fig.~\ref{fig3} the inverse AFM 
intraplane correlation lengths above $T_N^{\nu}$  are plotted, 
where the effects of the interplane
coupling and spin anisotropy are visible.
In the vicinity of $T_N^{+-}$ the temperature dependence of 
$(\xi_{xy}^{+-})^{-1}$ changes from an exponential law 
in the 2D case ($T_N^{+-}=0$) to a linear behavior for $R_z>0$.
Equally, $(\xi_{xy}^{zz})^{-1}$ near $T_N^{zz}$ behaves as
$T-T_N^{zz}$. According to the influence of spin anisotropy
on $T_N^{\nu}$ we get
$\partial \xi_{xy}^{zz}/\partial \Delta >0$ and  
$\partial \xi_{xy}^{+-}/\partial \Delta <0$. 
The behavior of $\xi_{xy}^{+-}$ qualitatively
agrees with the anisotropy dependence of the mean-field
expression for $\xi$ (being identified with $\xi_{xy}^{+-}$)
given in Ref.~\cite{Keea92b},
$\xi=\xi_0(1-\alpha_{eff}\xi_0^2)^{-1/2}$,
where $\xi_0$ denotes the correlation length
for $R_z=0$ and $\Delta=1$.

Finally let us consider the heuristic relation  between $T_N^{+-}$ 
and the transverse  2D correlation length at $T_N^{+-}$ 
which is often used in describing
the experimental data~\cite{CHN88} and is given by
$Q(R_z,\Delta =1)=0.25$ with 
$Q(R_z,\Delta)\equiv R_z(T_N^{+-})^{-1}[m^{+-}(T=0,R_z=0,\Delta)
\xi_{xy}^{+-}(T=T_N^{+-},R_z=0,\Delta)]^2$.
By our results, $Q(R_z,\Delta)$ in the experimentally relevant
region (cf. Sec.~IV) $2\times 10^{-4}\leq R_z\leq 2\times 10^{-2}$
and at $\Delta=1 (0.8)$ is found to vary between
0.21 (0.31) and 0.095 (0.23). That is, the heuristic
estimate is roughly confirmed by our theory.
\section{Comparison with experiments}
\begin{figure}[!b] 
\epsfig{file= 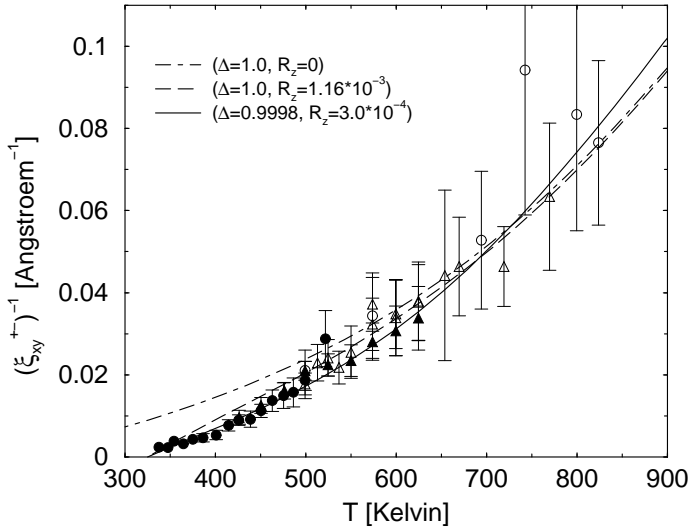, width =1.05 \linewidth}
\caption{Inverse antiferromagnetic transverse intraplane correlation
length in $\rm La_2CuO_4$ obtained by the neutron-scattering experiments 
of Ref.~\protect\cite{Biea99} (symbols) and from the theory for different
spatial and spin anisotropies.}
\label{fig4}
\end{figure}
Let us first compare our results for the transverse intraplane
correlation length $\xi_{xy}^{+-}$ with the neutron-scattering
data on $\rm La_2CuO_4$~(Ref.~\cite{Biea99}) in the range 
340~K$\leq T\leq$820~K plotted in Fig.~\ref{fig4}. Based on the
2D Heisenberg model ($\Delta=1$), in Ref.~\cite{WI98} the exchange 
energy $J$ was determined by a least-squares fit ($a$=3.79~{\AA}),
where for our choice of the vertex parameters the realistic value
$J=117$~meV was found. Here, we fix this value and consider the 
effects of spatial and spin anisotropy on $\xi_{xy}^{+-}(T)$. The
deviation of the theory for $R_z=0$ and $T<550$~K from the 
experimental data may be reduced by the inclusion of the interplane 
coupling, since $\xi_{xy}^{-1}(T_N)=0$. For $\Delta=1$ 
and $T_N=325$~K~\cite{Keea92a,Biea99} we obtain
$R_z=1.2\times 10^{-3}$, and the theoretical low-temperature
$\xi_{xy}^{-1}$ curve lies only somewhat above the experiments
(cf. Fig.~\ref{fig4}). Taking into account the spin anisotropy
$\alpha_{xy}=1.5\times 10^{-4}$~\cite{Keea92b} or  
$\alpha_{xy}=5.7\times 10^{-4}$~\cite{Biea99}, for $T_N^{+-}=325$~K
we get $R_z=3.0\times 10^{-4}$. For those parameters we obtain an
excellent  agreement between theory and experiment over the whole 
temperature region. Note that the theoretical curves for 
$\alpha_{xy}=2\times 10^{-4}$ (cf. Fig.~\ref{fig4}) and 
$\alpha_{xy}=2\times 10^{-3}$ agree within the accuracy of drawing.

Concerning our prediction of phase~II with longitudinal LRO 
in $\rm La_2CuO_4$,
for $\alpha_{xy}=1.5.\times 10^{-4}$  ($5.7\times 10^{-4}$) we obtain
the longitudinal zero-temperature magnetic moment 
$\mu^{zz}\equiv 2\mu_Bm^{zz}=6.6\times 10^{-2}\mu_B$
($6.1\times 10^{-2}\mu_B$) as compared with the transverse
moment $\mu^{+-}\equiv 2\mu_Bm^{+-}=0.55 \mu_B$.
 For the N\'{e}el temperature we 
find $T_N^{zz}=2.6\times 10^{-2}~J$ ($2.4\times 10^{-2}~J$)
equally to  $T_N^{zz}=35$~K (33~K) [$J=117$~meV].
With regard to the experimental verification of the two phases,
the magnitude of the longitudinal moment 
($\mu^{zz}\gtrsim 0.1\mu^{+-}$) may be large enough to 
allow a separation between $\mu^{zz}$ 
and $\mu^{+-}$ by polarized neutron-scattering studies on single crystals of
$\rm La_2CuO_4$~\cite{Kei01}. 

\begin{figure}[!h] 
\epsfig{file= 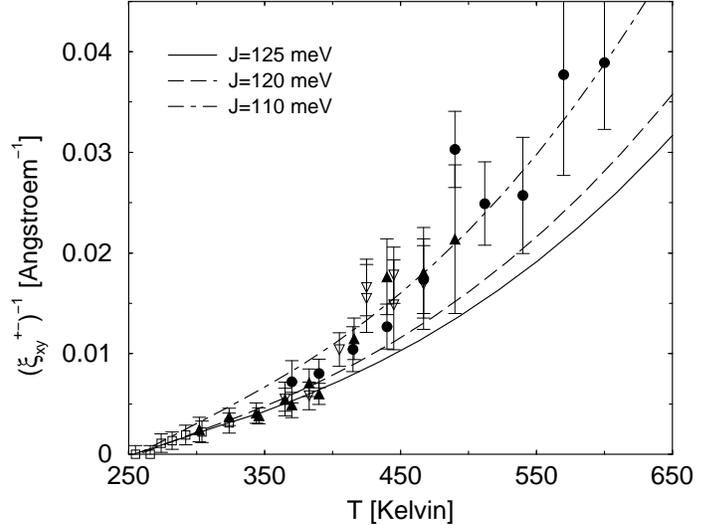, width =1.05 \linewidth}
\caption{Inverse antiferromagnetic transverse intraplane correlation
length in $\rm Sr_2CuO_2Cl_2$ obtained by the neutron-scattering experiments 
of Ref.~\protect\cite{Grea95} (symbols) and from the theory for different
exchange energies $J$.}
\label{fig5}
\end{figure}

Next we consider the compound 
$\rm Sr_2CuO_2Cl_2$ which is the best experimental realization of an $S=1/2$
2D Heisenberg antiferromagnet, where $J=125\pm 6$~meV, 
$\alpha_{xy}=1.4\times 10^{-4}$, and $T_N^{+-}=256.5$~K~\cite{Grea95}.
Keeping $T_N^{+-}$ and $\alpha_{xy}$ fixed (we take 
$\alpha_{xy}=2\times 10^{-4}$, as for $\rm La_2CuO_4$)
and choosing $J=125$~meV, 120~meV, and 110~meV, 
we obtain  the interplane coupling $R_z=2.0\times 10^{-5}$,
$4.0\times 10^{-5}$, and $1.0\times 10^{-4}$, respectively.
On the average  we have $R_z\simeq 5\times 10^{-5}\ll\alpha_{xy}$
in qualitative agreement with the estimate given in Ref.~\cite{Grea95}.
In Fig.~\ref{fig5}  our results for the transverse intraplane correlation 
length ($a=3.967$ {\AA}), where $(\xi_{xy}^{+-})^{-1}\propto T-T_N^{+-}$
near the transition to phase~I, are compared with the neutron-scattering
data~\cite{Grea95}. For $J=125$~meV we obtain a very good agreement
between theory and experiment at low enough temperatures ($T\lesssim 400$~K),
whereas for $J=110$~meV the agreement is good at higher temperatures.

The data on the predicted phase~II in $\rm Sr_2CuO_2Cl_2$  calculated for 
$J=125$~meV, 120~meV, and 110~meV are obtained as 
$\mu^{zz}/\mu_B=2.8\times 10^{-2}$, $3.9\times 10^{-2}$,
and  $5.5\times 10^{-2}$ (for comparison, $\mu^{+-}/\mu_B=0.54$)
and as $T_N^{zz}=7$~K, 12~K, and 18~K, respectively.
As in $\rm La_2CuO_4$, the magnitude of the longitudinal moment
($\mu^{zz}\simeq 4\times 10^{-2}\mu_B\simeq 0.08 \mu^{+-}$) may be
large enough to be detected by polarized neutron-scattering experiments 
on $\rm Sr_2CuO_2Cl_2$ single crystals.

Considering $\rm Ca_{0.85}Sr_{0.15}CuO_2$ 
($T_N=540$~K and $J=125$~meV~\cite{MRB99}),
for $\Delta=1$ we obtain $R_z=9.0\times10^{-3}$
as compared with $R_z\simeq 2.5\times10^{-2}$ resulting 
from a fit of the magnetization data~\cite{MRB99}.
Taking $\alpha_{xy}$ as for $\rm La_{2}CuO_4$,
 $\alpha_{xy}=1.5\times10^{-4}$~\cite{MRB99}, we get
$R_z=5.0\times10^{-3}$. For the zero-temperature
magnetic moments and the N\'{e}el temperature
we obtain $\mu^{zz}=0.16\mu_B$, $\mu^{+-}=0.57\mu_B$ 
and $T_N^{zz}=190~{\rm K}$, respectively.
Note that the longitudinal moment is more than twice as large
($\mu^{zz} = 0.28\mu^{+-}$) as in $\rm La_2CuO_4$. 
However, contrary to $\rm La_2CuO_4$ and $\rm Sr_2CuO_2Cl_2$, 
for $\rm Ca(Sr)CuO_2$ single crystals are not available, so that
neutron-scattering data do not exist until now~\cite{Kei01}.
\newpage
\section{Summary}
In this paper we presented a Green's-function theory for the quasi-2D
easy-plane XXZ model allowing the calculation of all static magnetic
properties at arbitrary temperatures, where we focused on the effects of
spatial and spin anisotropy on the AFM LRO and the correlation length. As a
qualitatively new result, for appropriate model parameters we obtained two
phase transitions, where the paramagnetic phase with pronounced AFM SRO
becomes unstable against a phase with transverse LRO only and, at a lower
temperature, a phase with both transverse and longitudinal LRO. Comparing
the theory with neutron-scattering experiments on the correlation length of
$\rm La_2CuO_4$, an excellent agreement is found. Furthermore, the second 
N\'{e}el transition (to the phase with longitudinal LRO)
in $\rm La_2CuO_4$, $\rm Sr_2CuO_2Cl_2$, and 
$\rm Ca_{0.85}Sr_{0.15}CuO_2$ is predicted 
to occur at about 30~K, 10~K, and 190~K, respectively. 
Our goal is to stimulate a wider discussion and new 
experiments in this direction.

{\it Acknowledgments.} 
The authors are greatly indebted to B. Keimer and R. Hayn
for stimulating discussions.
\section*{Appendix: Theory of Spin Susceptibility}
The spin susceptibilities 
$\chi^{+-}_{\bf q}(\omega)=
-\langle\langle S_{\bf q}^+;S_{- {\bf q}}^-\rangle\rangle_{\omega}$
and  $\chi^{zz}_{\bf q}(\omega)=-\langle\langle 
S_{\bf q}^z;S_{- {\bf q}}^z\rangle\rangle_{\omega}$,
($\langle\langle \dots;\dots\rangle\rangle_{\omega}$ denotes the
two-time retarded commutator Green's function)  
are determined by the projection method taking, as for the
XXZ chain~\cite{SFBI00}, the basis 
$(S_{\bf q}^+, i \dot{S}_{\bf q}^+)$ and $(S_{\bf q}^z, i 
\dot{S}_{\bf q}^z)$, respectively.  
We obtain
\begin{equation}
\label{chinu}
\chi^{\nu}_{\bf q}(\omega)=-\frac{M^{\nu}_{\bf q}}{\omega^2
-(\omega_{\bf q}^{\nu})^2}\,;\;\;\nu=+-,\,zz,
\end{equation}
with the first spectral moments 
$M_{{\bf q}}^{+-}=\langle [i\dot{S}_{\bf q}^+,S_{-{\bf q}}^-]\rangle$
and $M_{{\bf q}}^{zz}=\langle [i\dot{S}_{\bf q}^z,S_{-{\bf q}}^z]\rangle$
given by the exact expressions 
\begin{eqnarray}\label{mq1}
 M_{{\bf q}}^{+-}&=&-4 [C_{1,0,0}^{+-}(1-\Delta \gamma_{\bf q})
                      +2 C_{1,0,0}^{zz}(\Delta -\gamma_{\bf q})]
\\[0.1cm]
&&\hspace*{-0.1cm} - 2 R_z [C_{0,0,1}^{+-}(1-\Delta \cos q_z)
                      +2 C_{0,0,1}^{zz}(\Delta -\cos q_z)]\nonumber\\
M_{{\bf q}}^{zz}&=&-4 C_{1,0,0}^{+-}(1-\gamma_{\bf q}) - 2 R_z
C_{0,0,1}^{+-}(1-\cos q_z)\,,
\label{mq2}
\end{eqnarray}
$C_{nml}^{\nu}\equiv C_{\bf r}^{\nu}$, 
$C_{\bf r}^{+-}=\langle S_0^+ S_{\bf r}^- \rangle$,
$C_{\bf r}^{zz}=\langle S_0^z S_{\bf r}^z \rangle$,
${\bf r}= n {\bf e}_x +m  {\bf e}_y + l  {\bf e}_z $, and
$\gamma_{\bf q}=(\cos q_x +\cos q_y)/2$. 
The spin correlators are calculated as  
\begin{equation}
\label{cn}
C_{\bf r}^{\nu}=\frac{1}{N}\sum_{\bf q}\frac{M_{\bf q}^{\nu}}{2 
\omega_{\bf q}^{\nu}}
 \big[1+2p(\omega_{\bf q}^{\nu})\big] \mbox{e}^{i{\bf q r}}\,,
\end{equation}
where $p(\omega_{\bf q}^{\nu})=(\mbox{e}^{\omega_{\bf q}^{\nu}/T}-1)^{-1}$. 
The NN correlation functions are related to the internal energy per site
$\varepsilon=2(C_{1,0,0}^{+-}+\Delta C_{1,0,0}^{zz}) + R_z
(C_{0,0,1}^{+-}+\Delta C_{0,0,1}^{zz})$.

The spectra $\omega_{\bf q}^{\nu}$ are calculated in the approximations 
$-\ddot{S}^+_{\bf q}=(\omega_{\bf q}^{+-})^2 S^+_{\bf q}$ and 
$-\ddot{S}^{z}_{\bf q}=(\omega_{\bf q}^{zz})^2 S^z_{\bf q}$, where
products of three spin operators in $-\ddot{S}^+_{i}$ and
 $-\ddot{S}^z_{i}$ along NN sequences $\langle i,j,l\rangle$
are decoupled. Introducing vertex parameters in the spirit of the scheme
by Shimahara and Takada~\cite{ST91a} and extending the decouplings
given in Refs.~\cite{ISWF99,SIH00,SFBI00}, we have
\begin{eqnarray}
\label{d1}
S_i^+S_j^+S_l^-&=&\alpha^{+-}_{1x,1z} \langle S_j^+S_l^- \rangle S_i^+
+\alpha_2^{+-}   \langle S_i^+S_l^- \rangle S_j^+\,,\\\label{d2}
S_i^zS_j^+S_l^-&=&\alpha^{zz}_{1x,1z} \langle S_j^+S_l^- 
\rangle S_i^z\,,\\\label{d3}
S_i^+S_j^zS_l^-&=&\alpha^{zz}_{2} \langle S_i^+S_l^- \rangle S_j^z\,.
\end{eqnarray}
Here, $\alpha_{1x}^{\nu}$ and $\alpha_{1z}^{\nu}$ 
are attached to NN correlations in the $xy$-plane and along the $z$ direction, respectively, and $\alpha_2^{\nu}$ 
is associated with longer ranged correlation functions.
We obtain
\begin{eqnarray}
\label{w+-}
(\omega_{\bf q}^{+-})^2&=&
         [1 + 2 \alpha_2^{+-} (C_{2,0,0}^{+-}+2 C_{1,1,0}^{+-})
](1-\Delta \gamma_{\bf q})
\nonumber\\[0.1cm]
&&+\Delta [1 + 4 \alpha_2^{+-}(C_{2,0,0}^{zz}+
2 C_{1,1,0}^{zz})
](\Delta-\gamma_{\bf q})
\nonumber\\[0.1cm]
&&+2 \alpha_{1 x}^{+-} C_{1,0,0}^{+-}
[\Delta (4\gamma_{\bf q}^2 - 1)
- 3\gamma_{\bf q}]
\nonumber\\[0.1cm]
&&+4 \alpha_{1 x}^{+-} C_{1,0,0}^{zz}[4 \gamma_{\bf q}^2 - 1
- 3 \Delta\gamma_{\bf q}]
\nonumber\\[0.1cm]
&&+\frac{R_z^2}{2} \Big[(1 + 2 \alpha_{2}^{+-} C_{0,0,2}^{+-})
(1-\Delta\cos q_z)
\nonumber\\[0.1cm]
&&+\Delta (1 + 4 \alpha_{2}^{+-} C_{0,0,2}^{zz})
(\Delta-\cos q_z)
\nonumber\\[0.1cm]
&&+2 \alpha_{1 z}^{+-} C_{0,0,1}^{+-}
(\Delta \cos 2 q_z - \cos q_z)
\nonumber\\[0.1cm]
&&+4 \alpha_{1 z}^{+-} C_{0,0,1}^{zz}
(\cos 2 q_z - \Delta \cos q_z)\Big]
\nonumber\\[0.1cm]
&&2 R_z \Big[  2 \alpha_{2}^{+-} C_{1,0,1}^{+-}
(2 - \Delta \gamma_{\bf q} - \Delta\cos q_z)
\nonumber\\[0.1cm]
&&+4 \alpha_{2}^{+-} C_{1,0,1}^{zz}\Delta
(2\Delta - \gamma_{\bf q} - \cos q_z)
\nonumber\\[0.1cm]
&&+2 \alpha_{1 z}^{+-} C_{0,0,1}^{+-}
\gamma_{\bf q}(\Delta \cos q_z - 1)
\nonumber\\[0.1cm]
&&+4 \alpha_{1 z}^{+-} C_{0,0,1}^{zz}
\gamma_{\bf q}(\cos q_z - \Delta)
\nonumber\\[0.1cm]
&&+2 \alpha_{1 x}^{+-} C_{1,0,0}^{+-}
\cos q_z (\Delta\gamma_{\bf q}-1)
\nonumber\\[0.1cm]
&&\hspace*{0.4cm}+4 \alpha_{1 x}^{+-} C_{1,0,0}^{zz}
\cos q_z (\gamma_{\bf q}-\Delta)\Big]\,,\\[0.2cm]
\label{wzz}
(\omega_{\bf q}^{zz})^2&=&2(1-\gamma_{\bf q})
\Big[1 + 2 \alpha_{2}^{zz}(C_{2,0,0}^{+-}+2 C_{1,1,0}^{+-})
\nonumber\\[0.1cm]
&&\hspace*{0.4cm}-2\Delta \alpha_{1 x}^{zz} C_{1,0,0}^{+-} 
(1 + 4 \gamma_{\bf q})\Big]
\nonumber\\[0.1cm]
&&\hspace*{0.2cm}+R_z^2(1-\cos q_z) 
  \Big[1 + 2 \alpha_{2}^{zz} C_{0,0,2}^{+-}
\nonumber\\[0.1cm]
&&\hspace*{0.4cm}-2\Delta \alpha_{1 z}^{zz} C_{0,0,1}^{+-}
(1 + 2 \cos q_z ) \Big]
\nonumber\\[0.1cm]
&&\hspace*{0.2cm}+8 R_z\Big[ \alpha_{2}^{zz} C_{1,0,1}^{+-}
(2 - \gamma_{\bf q} - \cos q_z )
\nonumber\\[0.1cm]
&&\hspace*{0.4cm}+\Delta \alpha_{1 z}^{zz} C_{0,0,1}^{+-}
\gamma_{\bf q}(\cos q_z  - 1)
\nonumber\\[0.1cm]
&&\hspace*{0.4cm}+\Delta \alpha_{1 x}^{zz} C_{1,0,0}^{+-}
\cos q_z (\gamma_{\bf q} - 1)\Big]\,.
\end{eqnarray}
Note that in the special cases $R_z=0$ and $\Delta=1$ the spectra 
reduce to the expressions given in Refs.~\cite{FSWBI00} 
and~\cite{SIH00}, respectively.

The LRO in the correlators $C_{\bf r}^{\nu}$ is reflected in our theory by
the closure of the spectrum gap at ${\bf Q}=(\pi,\pi,\pi)$ as $T$ approaches
$T_N^{\nu}$ from above, so that 
$\lim_{T\to T_N^{\nu}}(\chi_{\bf Q}^{\nu})^{-1}=0$ and 
$\omega_{\bf Q}^{\nu}=0$ at $T\leq T_N^{\nu}$. 
Separating the condensation part $C^{\nu}\mbox{e}^{i{\bf Qr}}$ from
$C^{\nu}_{\bf r}$ [cf. Eq. (\ref{cn})], the magnetization $m^{\nu}$ 
is calculated as
\begin{equation}
\label{m}
(m^{\nu})^2=\frac{1}{N}\sum_{\bf r}C_{\bf r}^{\nu} \mbox{e}^{-i{\bf Q r}} 
=C^{\nu}\,.
\end{equation}

Considering the uniform static longitudinal susceptibility
$\chi_0^{zz}=\lim_{{\bf q}\to 0} M_{\bf q}^{zz}/(\omega_{\bf q}^{zz})^2$,
the ratio of the anisotropic functions $M_{\bf q}^{zz}$ and 
$(\omega_{\bf q}^{zz})^2=c_{xy}^2(q_x^2+q_y^2)+c_z^2q_z^2$ must be 
isotropic in the limit ${\bf q}\to 0$. This yields the condition
\begin{equation}
(c_{z}/c_{xy})^2=R_z C_{0,0,1}^{+-}/C_{1,0,0}^{+-}
\label{ellbed}
\end{equation}
with the squared spin-wave velocities
\begin{eqnarray}
\label{spiwexy}
c_{xy}^2&=&\frac{1}{2} + \alpha_2^{zz}(C_{2,0,0}^{+-}+2C_{1,1,0}^{+-})
                - 5\Delta\alpha_{1 x}^{zz} C_{1,0,0}^{+-} \nonumber
\\&&+2R_z(\alpha_2^{zz} C_{1,0,1}^{+-}
-\Delta\alpha_{1 x}^{zz} C_{1,0,0}^{+-})\,,\\[0.1cm]
c_{z}^2&=&R_z^2(\frac{1}{2} + \alpha_2^{zz}C_{0,0,2}^{+-}
                      - 3\Delta\,\alpha_{1 z}^{zz} C_{0,0,1}^{+-}) 
        \nonumber\\&& + 4R_z(\alpha_2^{zz} C_{1,0,1}^{+-}
-\Delta\alpha_{1 z}^{zz} C_{0,0,1}^{+-})\,.
\label{spiwezz}
\end{eqnarray}

Concerning the vertex parameters in our self-consistency scheme, 
three parameters are fixed by the sum rules 
$C_{0}^{+-}=1/2$, $C_{0}^{zz}=1/4$, and by Eq.~(\ref{ellbed})
for all $T$.
To determine the free parameters taken as $\alpha_{1x,1z}^{+-}$
and $\alpha_{2}^{zz}$, we need additional conditions.
Let us consider the ground-state energy per site which we
compose approximately, following Ref.~\cite{SIH00}, as
$\varepsilon(\Delta,R_z)=e(\Delta,R_z)+e(\Delta,1)-e(\Delta,0)$.
Here, $e(\Delta,R_z)$ denotes the ground-state energy of the 2D
spatially anisotropic XXZ model [Eq.~(\ref{xxzmod}) without sum 
in $y$ direction], where the values in the 1D ($R_z=0$) and 2D
cases ($R_z=1$) are taken from Ref.~\cite{YY66a} and the 
exact data of Refs.~\cite{FSWBI00,OK88}, 
respectively. Since $e(\Delta,R_z)$ is known
for $R_z=0,1$ and $\Delta=1$ (taken from the Ising-expansion results
by Affleck et al.~\cite{AGS94}), we approximate $e(\Delta,R_z)$ by the 
linear interpolation 
$e(\Delta,R_z)=e(\Delta,0)+[e(\Delta,1)-e(\Delta,0)]
[e(1,1)-e(1,0)]^{-1}[e(1,R_z)-e(1,0)]$.
At $T=0$, we adjust  $\alpha_{1x}^{+-}$ to  $\varepsilon(\Delta,R_z)$
and $\alpha_{2}^{zz}$ to $\partial \varepsilon/\partial \Delta
=2 C_{1,0,0}^{zz} + R_z C_{0,0,1}^{zz}$. 

To formulate conditions for $\alpha_{1x}^{+-}$ and 
$\alpha_{2}^{zz}$ also at finite temperatures, we follow
the reasonings of Refs.~\cite{ST91a,WI97,SFBI00}. 
That means, we conjecture that the ``vertex corrections'' 
$\alpha_{1x}^{\nu}(T)-1$ and $\alpha_{2}^{zz}(T)-1$ have similar  
temperature dependences and vanish in the high-$T$ limit. 
Correspondingly, as the simplest interpolation between 
high temperatures and $T=0$ we assume the ratio of two 
vertex corrections as temperature independent and fixed
by the ground-state value, i.e.,  
\begin{eqnarray}
\label{r1}
\frac{\alpha_{1x}^{+-}(T)-1}{\alpha_{1x}^{zz}(T)-1}&=& {\rm const.}\\[0.1cm]
\frac{\alpha_{2}^{zz}(T)-1}{\alpha_{1x}^{zz}(T)-1}&=& {\rm const.}
\label{r2}
\end{eqnarray}
To determine $\alpha_{1z}^{+-}(T)$, as compared 
with $\alpha_{1z}^{zz}(T)$ fixed by the ``isotropy condition''~(\ref{ellbed})
resulting from $\chi_0^{zz}$, we first note that an analogous condition
cannot be derived from 
$\chi_0^{+-}=\lim_{{\bf q}\to 0} M_{\bf q}^{+-}/(\omega_{\bf q}^{+-})^2$,
since both $M_{\bf q}^{(1)}$ and $\omega_{\bf q}^{+-}$ have non-zero
${\bf q}\to 0$ limits [cf. Eqs.~(\ref{mq1}) and~(\ref{w+-})]. Therefore,
for $\alpha_{1z}^{+-}(T)$ we make the plausible ansatz assuming the ratio
$\alpha_{1z}^{\nu}(T)/\alpha_{1x}^{\nu}(T)$ as $\nu$ independent, i.e., 
\begin{equation}
\label{r3}
\frac{\alpha_{1z}^{+-}(T)}{\alpha_{1x}^{+-}(T)}= 
\frac{\alpha_{1z}^{zz}(T)}{\alpha_{1x}^{zz}(T)}\,.
\end{equation} 

From the solution of the self-consistency equations the AFM correlation
lengths above $T_N^{\nu}$ may be evaluated. They are obtained by 
the expansion of $\chi_{\bf q}^{\nu}$ around ${\bf Q}$,
$\chi_{\bf q}^{\nu}=\chi_{\bf Q}^{\nu}[1+(\xi_{xy}^{\nu})^2(k_x^2+k_y^2)
+(\xi_{z}^{\nu})^2 k_z^2]^{-1}$ with ${\bf k}={\bf q}-{\bf Q}$. 
The squared intraplane correlation lengths are given by 
\begin{eqnarray}
\label{xixy}
(\xi_{xy}^{+-})^2&=&-(\omega_{\bf Q}^{+-})^{-2}
\\&&\hspace*{-0.8cm}
\times\Big[\mbox{\small $\frac{\Delta}{2}$}
+\Delta\alpha_2^{+-}\big(\mbox{\small $\frac{1}{2}$}
C_{2,0,0}^{+-}+ C_{2,0,0}^{zz} 
+  C_{1,1,0}^{+-} + 2 C_{1,1,0}^{zz}\big)
\nonumber\\&&\hspace*{-0.8cm}
+ \alpha_{1 x}^{+-}\big((4 \Delta +\mbox{\small $\frac{3}{2}$}) C_{1,0,0}^{+-}
+ (8 + 3 \Delta) C_{1,0,0}^{zz}\big)
\nonumber\\&&\hspace*{-0.8cm}
+ R_z\big[\alpha_{1 x}^{+-}\big(\Delta C_{1,0,0}^{+-}+2 C_{1,0,0}^{zz}\big) 
+  \alpha_{1 z}^{+-}
\big(C_{0,0,1}^{+-}+2 C_{0,0,1}^{zz}\big)
\nonumber\\&&\hspace*{-0.8cm}
+ \Delta\big( \alpha_{1 z}^{+-}
(C_{0,0,1}^{+-}+2 C_{0,0,1}^{zz})+\alpha_2^{+-}
( C_{1,0,1}^{+-}
+2 C_{1,0,1}^{zz})\big)\big]\Big]
\nonumber\\&&\hspace*{-0.8cm} 
-\big(\Delta C_{1,0,0}^{+-}+2 C_{1,0,0}^{zz}\big)
/M_{\bf Q}^{+-}\,,
\nonumber\\[0.1cm]
(\xi_{xy}^{zz})^2&=&-(\omega_{\bf Q}^{zz})^{-2}
\\&&\hspace*{-0.8cm}
\times\Big[
\mbox{\small $\frac{1}{2}$} 
+ \alpha_2^{zz} \big(C_{2,0,0}^{+-}+2 C_{1,1,0}^{+-}\big)
+ 11 \Delta\alpha_{1 x}^{zz} C_{1,0,0}^{+-}
\nonumber\\&&\hspace*{-0.8cm}
+ 2 R_z\big(\alpha_{1 x}^{zz} \Delta C_{1,0,0}^{+-} + 
  2 \alpha_{1 z}^{zz}\Delta C_{0,0,1}^{+-} 
+ \alpha_2^{zz} C_{1,0,1}^{+-}\big)\Big]\nonumber\\&&\hspace*{-0.8cm} 
-2 C_{1,0,0}^{+-}/M_{\bf Q}^{zz}\,.
\nonumber
\label{xizz}
\end{eqnarray}

\end{document}